\documentclass[%
 aps,
 prl,
 amsmath,amssymb,
 reprint,%
]{revtex4-2}

\usepackage{graphicx}
\usepackage{dcolumn}
\usepackage{bm}
\usepackage{caption}
\usepackage{subcaption}
\usepackage[utf8]{inputenc}
\usepackage[T1]{fontenc}
\usepackage{mathptmx}
\usepackage{tabularx}
\usepackage{graphicx}
\usepackage{multirow}
\usepackage{siunitx,color}
\usepackage{booktabs}

\begin{document}

\title{Absence of Space-Charge-Limited Current in Unconventional Field Emission}

\author{Cherq Chua}
\affiliation{Science, Mathematics and Technology (SMT), Singapore University of Technology and Design, 8 Somapah Road, Singapore 487372}

\author{Chun Yun Kee}
\affiliation{Science, Mathematics and Technology (SMT), Singapore University of Technology and Design, 8 Somapah Road, Singapore 487372}

\author{Yee Sin Ang}
\email{yeesin\_ang@sutd.edu.sg}
\affiliation{Science, Mathematics and Technology (SMT), Singapore University of Technology and Design, 8 Somapah Road, Singapore 487372}

\author{L. K. Ang}
\email{ricky\_ang@sutd.edu.sg}
\affiliation{Science, Mathematics and Technology (SMT), Singapore University of Technology and Design, 8 Somapah Road, Singapore 487372}

\begin{abstract}
For field emission (FE), it is widely expected that its emitting current density $J$ will become space-charge-limited current (SCLC) due the built-up of charge in-transit within a gap spacing $D$ biased at sufficiently large voltage $V$.
In this paper, we reveal a peculiar finding in which this expected two-stage transition (from FE to SCLC) is no longer valid for FE not obeying the traditional Fowler-Nordheim (FN) law. 
By employing a generalized FN scaling of $\ln\left(J/V^k\right) \propto - 1/V$, we show the existence of a \emph{critical exponent} $k_c \equiv 3/2$ where unusual behaviours occur for $k < k_c$:
(a) Only FE at small $D$ (no transition to SCLC even at infinitely large $V$), and 
(b) Three-stage transition from FE first to SCLC then back to FE at large $D$.
For any $k > k_c$, the conventional two-stage transition from FE to SCLC will always occur for all $D$, which also includes the conventional FN law at $k$ = 2.
Using various unconventional FE models with $k \neq 2$, we specifically demonstrate these peculiar transitions.
Under a normalized model, our findings uncover the rich interplay between the source-limited FE and bulk-limited SCLC over a wide range of operating conditions.
\end{abstract}

\maketitle

\textcolor{blue}{\textbf{\emph{Introduction.}}} Field emission (FE) describes the process of electron emission from a cathode to free space via the quantum mechanical tunneling at a sufficiently high electric field.
It is known as the Fowler-Nordheim (FN) law
\cite{FN1928} derived decades ago:

%
\begin{equation}\label{eqn:FN}
    J = \mathcal{A} V^2 \exp\left( - \frac{\mathcal{B}}{ V } \right),
\end{equation}
where $J$ is the current density, $V$ is the bias voltage, $\mathcal{A}$ and $\mathcal{B}$ are material-and structure-dependent parameters. 
At high $V$ regime where $J$ reaches a sufficiently large quantity that the space charge effects become important, the FN law transits into the space-charge-limited current (SCLC). 
The simplest model of SCLC is governed by the one-dimensional (1D) Child-Langmuir (CL) law \cite{PhysRevSeriesI.32.492,PhysRev.2.450}:
%
\begin{equation} \label{eqn:CL}
    J = \frac{4\epsilon_0}{9} \sqrt{\frac{2e}{m_e}} \frac{V^{3/2}}{D^2},  
\end{equation}
where $e$ and $m_e$ is the charge and the mass of free electron, respectively, $\epsilon_0$ is the free space permittivity, $V$ is the gap voltage, and $D$ is the gap spacing.
For a nanogap where quantum effect is important, the quantum CL law leads to a distinctive scaling of $J\propto V^{1/2}$ \cite{quantumSCLC}. 
The transition from the FN law to CL law had been studied for a large gap \cite{doi:10.1063/1.870603,doi:10.1063/1.2226977,doi:10.1063/1.3272690}, micro-gap \cite{doi:10.1063/1.4914855} and nanogap \cite{doi:10.1063/1.2378405,Koh_2008}.
The transition from source-limited FE to high-current SCLC is important to understand the current transport in vacuum diodes \cite{doi:10.1063/1.4978231} and nanodiodes \cite{doi:10.1063/5.0042355}. 


The conventional two-stage transition from FE (at low $V$) to SCLC (at high $V$) is based on the assumption that the FE follows the classical FN law [see Eq. (\ref{eqn:FN})] having a scaling of $\ln\left(J/V^k\right) \propto V^{-1}$ with $k$ = 2. 
This understanding seems to be challenged by the emergence of various
materials and settings not obeying the classical emission models
\cite{MRS2017,IM2021,doi:10.1063/1.3140602,doi:10.1063/5.0047771,
ang2020universal,weijie2021}.
Examples of FE model with $k\neq$2 include Schottky-Nordheim (SN) barrier modification \cite{doi:10.1063/1.3140602} with $k \approx 1.23$ for a work function of 4.5 eV, graphene with $k$ = 1.53 \cite{doi:10.1063/5.0047771}, 2D semimetals with $k$ = 1  \cite{ang2020universal}, and 3D topological semimetals with $k$ = 3 \cite{weijie2021}.
Note $k \neq 2$ behavior of FE can also be caused by geometrical nature of the surface, such as $k<$ 2 for a rough surface \cite{Zubair_2018} and its corresponding SCLC can be enhanced beyond the 1D CL law \cite{article3}.



\begin{figure*}[]
    \centering
    \includegraphics[width=15.5cm]{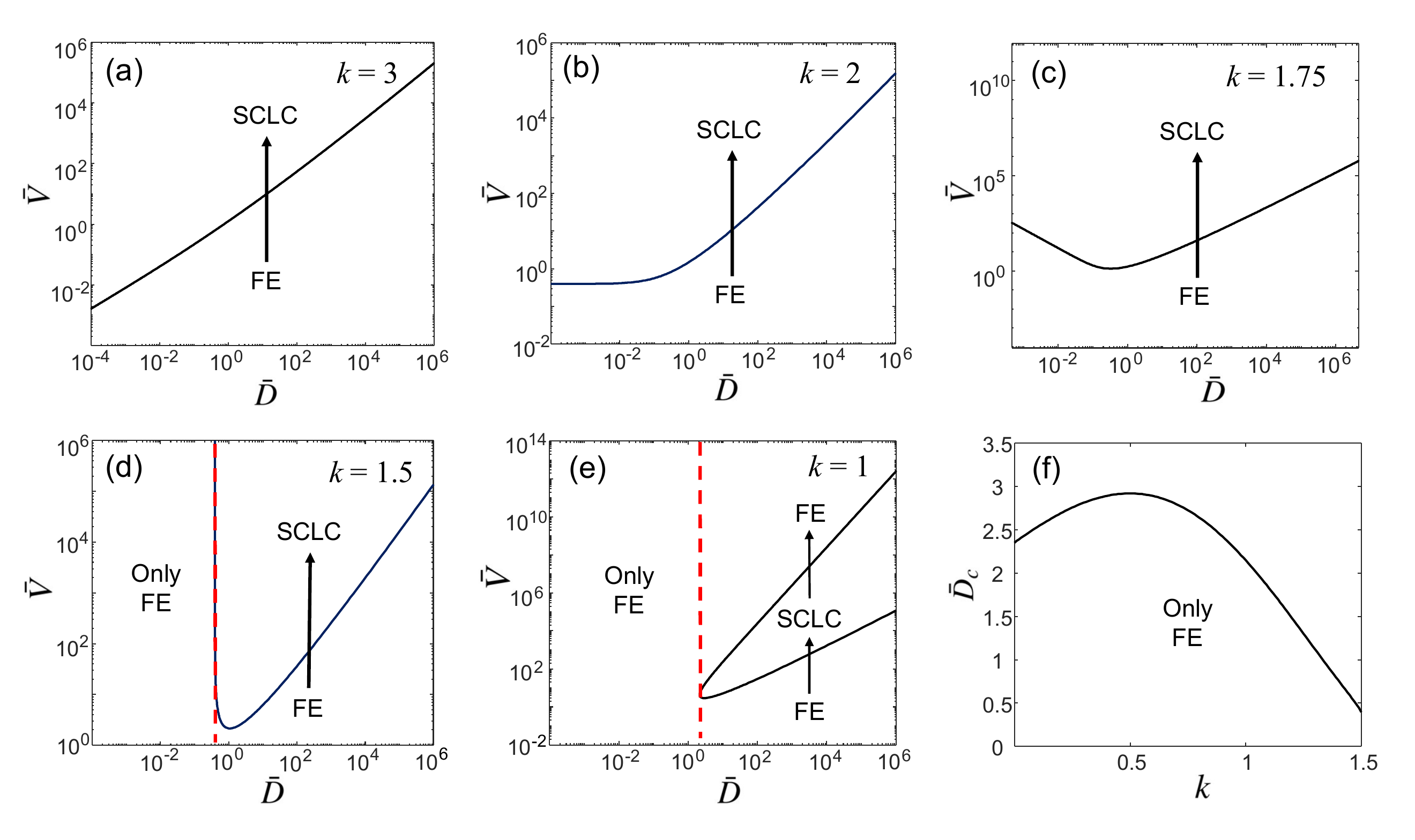}
    \caption{Susceptibility diagram of field emission (FE) and space-charge-limited current (SCLC) for normalized voltage $\bar{V}$ 
    and normalized gap spacing $\bar{D}$. The solid line governs the transition between FE and SCLC for (a) $k$ = 3, (b) $k$ = 2, (c) $k$ = 1.75, (d) $k = k_c \equiv 3/2$, and (e) $k$ = 1. The red dashed line in (d) and (e) indicates the critical $\bar{D_c}$ which separates the exclusively only FE from the multiple-stage transition. (f) The critical $\bar{D_c}$ from $k$ = 0 to $k_c \equiv 3/2$.}
    \label{fig:Fig 1}
\end{figure*}

In this paper, we aim to study the transition from FE to SCLC for arbitrary values of $k$ over a wide range of applied voltage $V$ and gap spacing $D$.
The main findings are summarized in Fig. \ref{fig:Fig 1} and Table I.
Intriguingly, there is a critical value $k_c \equiv 3/2$ that, for any $k \leq k_c$, the transition to SCLC is no longer possible for small gap spacing $D$ [see Fig. \ref{fig:Fig 1}(d) and Fig. \ref{fig:Fig 1}(e)].
For large gap spacing $D$ at $k < k_c$, the transition becomes a three-stage process: first from FE to SCLC and then back to FE as shown in Fig. \ref{fig:Fig 1}(e).
For $k > k_c$, we have two-stage transition from FE to SCLC for all values of $D$ [see Figs. \ref{fig:Fig 1}(a) to \ref{fig:Fig 1}(c)].
Here Fig. \ref{fig:Fig 1}(b) at $k$ = 2 is the usual transition from the 1D FN law to the 1D CL law \cite{doi:10.1063/1.870603}.

The above mentioned abnormal transition (no transition to SCLC) arises solely because of unconventional FE having $k < k_c$, which fails to supply sufficient space charge to sustain the required SCLC imposed by the 1D CL law.
The critical $k_c \equiv 3/2$ can be qualitatively explained by the $3/2$-scaling of the surface electric field governed by the 1D classical CL law \cite{PRL2013}. 
It is not due to various enhancements of the CL law such as quantum effects \cite{quantumSCLC,PhysRevLett.66.1446}, 
finite emission area \cite{PRL1996john,PRL2001lau}, sharp tips \cite{doi:10.1063/1.4919936}, 
short-pulse \cite{doi:10.1063/1.1463065,PRL2007Ang,Wu2008}
and
finite particles \cite{PRL2010,Zhu2011}. 
The breakdown of materials at high field is also ignored completely.

Our generalized model shall provide an important foundation for the modeling of unconventional FE and its transition to high current regime required for applications in high power microwave sources \cite{doi:10.1063/1.2838240} and vacuum electronics \cite{5993476,PRL2005Edl,PRL2005temkin}. 
While the model is derived only for a planar gap, we expect the critical $k_c \equiv 3/2$ is applicable to cylindrical and spherical gap, as these geometries have been shown to have the same scaling of the 1D CL law \cite{PRL2013}.



%


\textcolor{blue}{\textbf{\emph{Model.}}} We consider a generalized FE current density of   
\begin{equation}
    J = \mathcal{A}_k {E_s}^k \exp\left( - \frac{\mathcal{B}}{E_s}\right),
\end{equation}
where $k$ is an arbitrary \emph{scaling} of the surface electric field $E_s$ in the pre-exponent term, $\mathcal{A}_k$ and $\mathcal{B}$ are constants due to materials and other properties. 
Note Eq. (3) recovers the classic 1D FN law at $k$ = 2. 
For simplicity, Eq. (3) can be transformed into a normalized form of
\begin{equation}
    \bar{J} = \bar{E}^k \exp\left( -\frac{1}{\bar{E}} \right) ,
\end{equation}
with $\bar{J} = J/J_{0}$, $\bar{E} = E_s / E_0$.
The normalized scales are $E_0 = \mathcal{B}$ (electric field) and $J_{0} = \mathcal{A}_k \mathcal{B}^k$ (current density).
 
Using the same approach \cite{doi:10.1063/1.870603} in linking Eq. (4) to the Poisson equation in the Llewellyn form to account for the space charge effects, a set of governing equations is obtained:

%
%
%
%
%

%
\begin{equation}
    \frac{\bar{J} \bar{T}^2}{2} + \bar{E} \bar{T} = \sqrt{2\bar{V}} ,
\end{equation}
\begin{equation}
    \frac{\bar{J}\bar{T}^3}{6} + \frac{\bar{E}\bar{T}^2}{2} = \bar{D} .
\end{equation}
Here the other normalized variables are 
gap spacing $\bar{D} = D/D_0$, 
transit time $\bar{T} = T/T_0$, and 
applied voltage $\bar{V} = V /V_0$.
Their respective scales are 
$D_0 = eE_0{T_0}^2/m_e$ (length), 
$T_0 = \varepsilon_0E_0/J_{0}$ (time), and 
$V_0 = E_0 \times D_0$ (voltage). 
Once an emission model (or material) is specified with known $\mathcal{A}_k$ and $\mathcal{B}$, all the normalized constants can be calculated. 
Solving Eq. (5), we obtain
$\bar{T} = \left({\xi_k}/{\bar{E}^{k-1}}\right) \exp\left( 1/\bar{E} \right)$, and
\begin{equation}
    \xi_k =  -1 + \sqrt{ 1 + 2\left(2\bar{V}\right)^{1/2} \bar{E}^{k-2} \exp\left( -\frac{1}{\bar{E}} \right)} , 
\end{equation}
which will allow Eq. (6) to be rewritten as
\begin{equation}
    {\xi_k}^3 + 3{\xi_k}^2 = 6\bar{D}\bar{E}^{2k-3} \exp\left( - \frac{2}{\bar{E}} \right) .
\end{equation}
\begin{figure*}[]
    \centering
    \includegraphics[width=16cm]{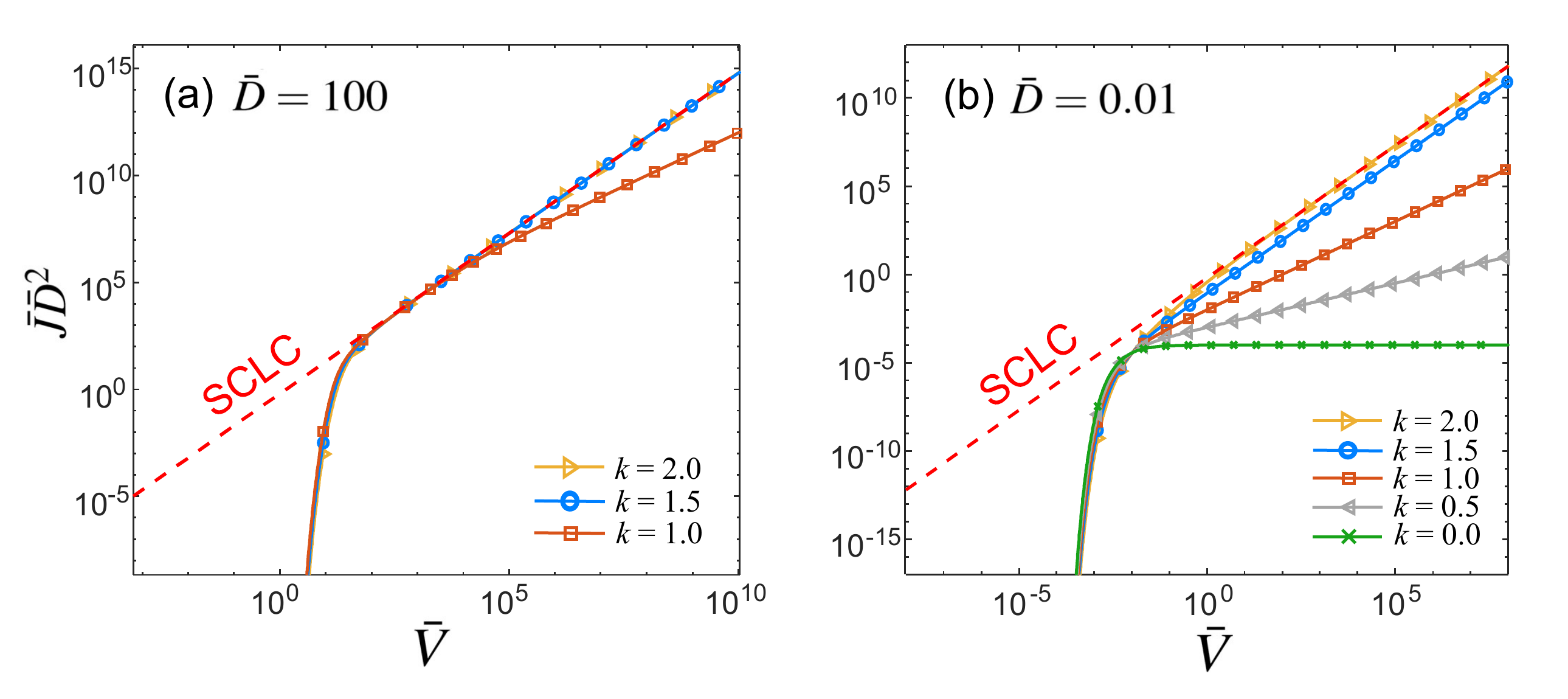}
    \caption{Normalised $\bar{J}$ - $\bar{V}$ characteristics of field emission with different $k$ at (a) $\bar{D}=100$, and (b) $\bar{D}=0.01$. The red dashed lines represent the SCLC limit.}
    \label{fig:Fig 2}
\end{figure*}
For given $k$ and $\bar{D}$, Eqs. (7) and (8) are solved numerically to obtain $\bar{V}$ as a function of $\bar{E}$. 
Using this solved $\bar{V}(\bar{E})$, we calculate 
$\bar{J}$ via Eq. (4), and 
the relationship between $\bar{J}$, $\bar{D}$ and $\bar{V}$ is determined consistently for arbitrary $k$.

The factor $\bar{E}^{2k-3}$ in Eq. (8) suggests two contrasting behaviors as $\bar{E} \gg$ 1.
For $2k>3$, we have
$6\bar{D}\bar{E}^{2k-3} \exp\left(-2/\bar{E}\right) \to 6\bar{D}\bar{E}^{2k-3}$.
For $2k<3$, the limit is $6\bar{D}\bar{E}^{2k-3} \exp\left(-2/\bar{E}\right) \to 0$.
Such contrasting limit of Eq. (8) at $\bar{E} \gg$ 1 indicates
the existence of a \emph{critical} value of $k$ defined as 
$k_c \equiv 3/2$, which classifies into three different regimes: 
(i) $k > k_c$ (super-critical),
(ii) $k = k_c$ (critical) and 
(iii) $k < k_c$ (sub-critical).

In the super-critical regime at $k > k_c \equiv 3/2$  , the normalized FE model at low field $\bar{E} \ll 1$ is
\begin{equation}
    \bar{J}_{FE} = \left(\frac{\bar{V}}{\bar{D}}\right)^k \exp\left(-\frac{\bar{D}}{\bar{V}}\right).
\end{equation}
At high field $\bar{E} \gg 1$, we recover the normalized 1D CL law:  
\begin{equation}
    \bar{J}_{SCLC} \times \bar{D}^2 = \frac{4\sqrt{2}}{9} \times \bar{V}^{3/2}.
\end{equation}
Detailed derivation for Eqs. (9) and (10) and the analytical formulation of asymptotic limits at the regimes can be found in the Supplementary Materials \cite{supplementalmaterial}.

\begin{table}[b]
\caption{Transition behaviours at different $k$ and normalized gap spacing $\bar{D}$. Note $k_c \equiv 3/2$ and $\bar{D_c}$ from Fig. 1(f)}.
\begin{ruledtabular}
\begin{tabular}{ccc}
$k$              & $\bar{D}$ & Transition behaviour                   \\
\midrule
$k>k_c$               & All $\bar{D}$             &  2-stage: FE-to-SCLC         \\
\midrule
$k=k_c$           & $\bar{D} < \bar{D_c}$                    & Only FE          \\  
                                 & $\bar{D} > \bar{D_c} $  
                                 & 2-stage: FE-to-SCLC                          \\ 
 \midrule
$k<k_c$ & $\bar{D} < \bar{D_c}$                    & Only FE \\  
                                 & $\bar{D} > \bar{D_c}$                    & 3-stage: FE-to-SCLC-to-FE                      \\ 
\end{tabular}
\end{ruledtabular}
\end{table}

\textcolor{blue}{\textbf{\emph{Results \& Discussion.}}} 
By equating Eqs. (9) and (10), we obtain the relationship between $\bar{V}$ and $\bar{D}$, which is
\begin{equation}
    \frac{4\sqrt{2}}{9}\frac{\bar{V}^{3/2-k}}{\bar{D}^{2-k}} = \exp\left(-\frac{\bar{D}}{\bar{V}}\right).
\end{equation}
Solutions of Eq. (11) are plotted in Fig. 1 for different $k$ = 1, 1.5, 1.75, 2 and 3. to show the boundary separating the FE-dominated and SCLC regimes as a function of $\bar{V}$ and $\bar{D}$. 

For super-critical cases with $k > k_c \equiv$ 1.5 [such as $k$ = 3, 2 and 1.75 in Fig. 1(a), 1(b) and 1(c)], the 2-stage FE-to-SCLC transition is evident at increasing $\bar{V}$ with all $\bar{D}$. 
Such classic transition is, however, no longer warranted for $k \leq 1.5$.
For critical ($k=k_c\equiv 3/2$) and sub-critical ($k=1 < k_c$) cases, we have only FE regime (complete absence of SCLC even at infinitely large $V$) when $\bar{D}$ is smaller than a critical value $\bar{D_c}$ [marked by the red dashed lines in Figs. 1(d) and 1(e)]. 
This critical $\bar{D_c}$ is defined as the minimum solution of $\bar{D}$ from Eq. (11) as plotted in Fig. 1(f) for $k$ = 0 to 1.5.
For $\bar{D} > \bar{D_c}$, the two-stage transition (FE-to-SCLC) is only present in the critical case ($k=1.5$) in Fig. 1(d). 
This is in stark contrast to the sub-critical case ($k=$ 1) in Fig. 1(e) for which the SCLC occurs only in the intermediate $\bar{V}$ regime before the FE becomes dominant again at high $\bar{V}$, and it can be viewed as the three-stage transition (FE-to-SCLC-to-FE).

\begin{figure*}
    \centering
    \includegraphics[width=17cm]{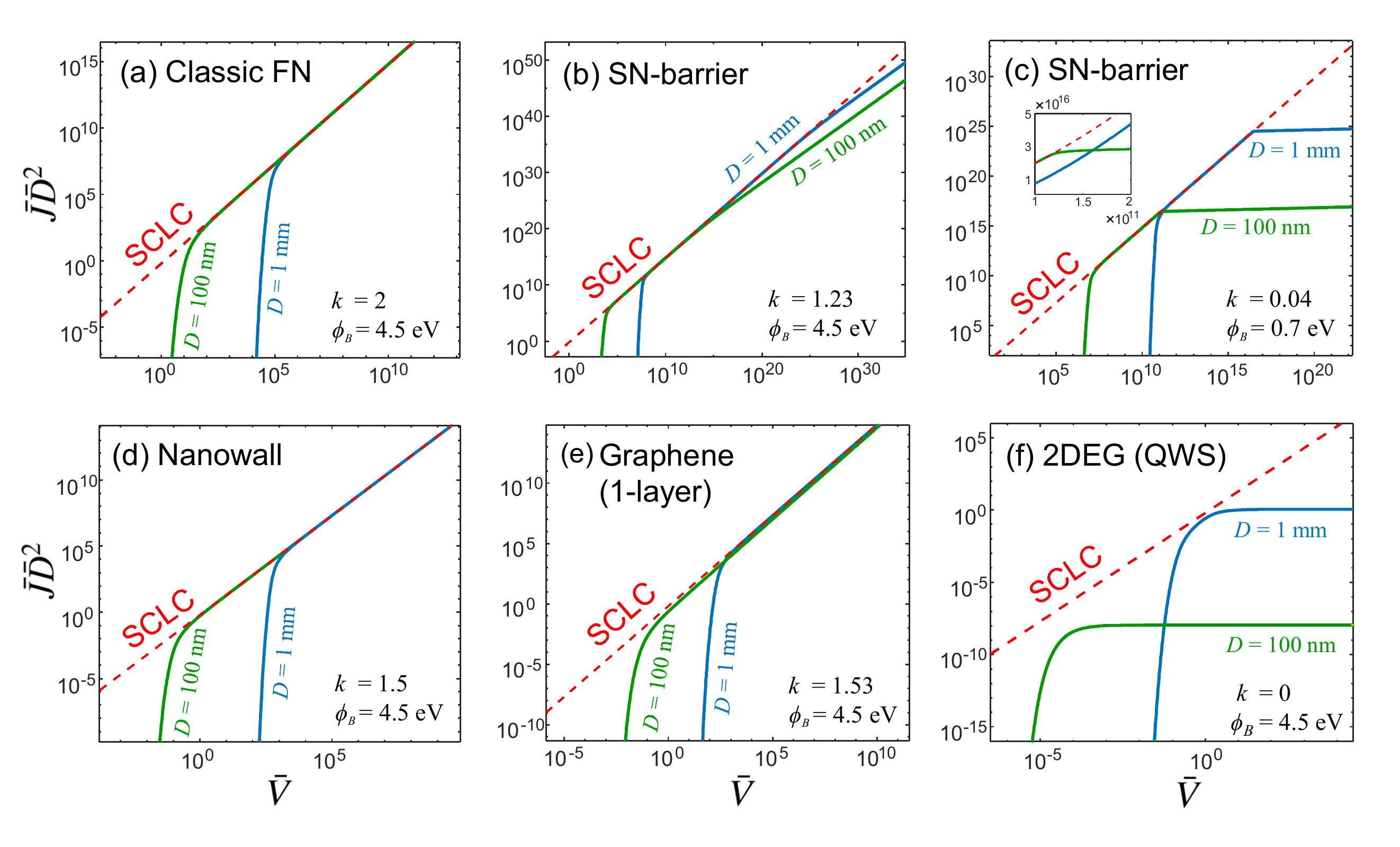}
    \caption{$\bar{J}$ - $\bar{V}$ curves of different field emission models at $D=100$ \si{nm} and 1 \si{mm}. The work functions $\phi_B$ used for each model are labeled in the figures. The red dashed lines represent the SCLC limits.}
    \label{fig:Fig 3}
\end{figure*}
%
%
Fig. 2 shows the calculated values of $\bar{J} \times \bar{D}^2$ as a function of $\bar{V}$ of various $k$ at (a) large $\bar{D}$ = 100 and (b) small $\bar{D}$ = 0.01.
For $k$ = 2 (super-critical cases), the results exhibit a smooth transition from FE to SCLC for both large and small $\bar{D}$ cases, which confirms Fig. 1(b) scenario and thus also recovers the well-known transition from the 1D FN law to the 1D CL law \cite{doi:10.1063/1.870603}.
As discussed above, this two-stage transition is no longer warranted for $k \leq 1.5$, and $\bar{D}$ will become an important factor in determining the transition.
For small $\bar{D}$ = 0.01, Fig. 2(b) shows that transition from FE to SCLC is not possible for both $k$ = 1.5 and $k$ = 1, which confirm the results in Figs. 1(d) and 1(e).
On the other hand, Fig. 2(a) confirms the three-stage transition only occurs at large $\bar{D}$ = 100 for $k=1$ case, which is also depicted in Fig. 1(e). 
The three-stage transition arises because FE could only generates sufficient current density to maintain SCLC at moderate voltage, and thus back to FE at high voltage.

The above different transitions for various $k$ and $D$ are summarized qualitatively in Table I.
The \emph{exclusive and only} FE mechanism at $k \leq k_c \equiv 1.5$ and at small $\bar{D} < \bar{D_c}$ arises due to not having sufficient emitted electrons to sustain the SCLC in a small gap spacing. 
Fig. 2(b) also shows a significant reduced current density with decreasing 
$k$ towards zero and saturation of $J$ at $k$ = 0 limit.
If one can identify some FE models with $k \approx$ 0, this property may serve a favorable operating condition for stable current output even at high voltage [see Fig. 3(c) and 3(f) below].



We now employ several unconventional FE models at different $k$ as examples to illustrate the peculiar transition reported above.
The details of each model are discussed and their normalized constants are calculated \cite{supplementalmaterial}.
For simplicity, we consider work function $\phi_B$ = 4.5 \si{eV} for 5 cases: 
classical FN law ($k$ = 2), Schottky-Nordheim (SN) barrier ($k$ = 1.23), nanowall ($k$ = 1.5), mono-layer graphene ($k$ = 1.53) and two-dimensional electron gas (2DEG) ($k$ = 0) as shown in Fig. 3(a), 3(b), 3(d) 3(e) and 3(f), respectively.
For each model, we have used large and small gap spacing: $D$ = 1 mm and 100 nm.
As an example, for $D$ = 100 \si{nm}, the FE-SCLC transition is at $\bar{V}$ = 26.47 in Fig.3(a), which corresponds to about $V \approx 3 \times {10}^3$ \si{V} or electric field $E_s \approx$ 30 \si{V/nm}. 
For $D$ = 1 \si{mm}, it is about $E \approx$ 10 \si{V/nm}.
This high field in transition can be reduced by using lower work function [see Fig. 3(c)].

Instead of the simple triangular barrier used in the FN law, the SN barrier model \cite{doi:10.1063/1.3140602} has included the image charge effect, and the $k$ is calculated by $k=2-\eta/6$,
where $\eta \equiv e^3 B_{FN}/4\pi\epsilon_0{\phi_B}^{1/2}$ is a dimensionless work-function-dependent parameter.
For $\phi_B$ = 4.5 \si{eV}, we have $k \approx 1.23$ used in Fig. 3(b) and it has a lower transition voltage as compared to $k$ = 2 in Fig. 3(a).
The lower-field at transition is expected, as the presence of image charge effect could reduce the tunneling barrier.
Note Fig. 3(b) belongs to the sub-critical case, where the emission will eventually dwells back into the FE regime at high voltage due to the three-stage transition.
However, the transition from SCLC back to FE will occur at an unrealistically high voltage, which is not feasible.

For critical case at $k \approx k_c \equiv$ 1.5, we consider nanowall ($k$ = 1.5) and graphene ($k$= 1.53), which both converge to SCLC for both $D$ = 100 nm and $D$ = 1 mm.
The graphene case is based on a Bardeen transfer Hamiltonian formalism \cite{doi:10.1063/5.0047771}. 
The nanowall case is due to the quantum confinement effect leading to a quantized energy component along the confinement direction \cite{Qin_2010}, which gives $k \approx 1.5$. 

%
%

%
%
As mentioned before, transition from FE to SCLC at low field is feasible by using low work function materials \cite{doi:10.1021/acsenergylett.9b01214,jacobs}.
For $\phi_B$ = 0.7 \si{eV} \cite{doi:10.1021/acsenergylett.9b01214}, SN model gives $k$ = 0.04 used in Fig. 3(c), which shows a low transition E-field of about 0.3 \si{V/nm} for both $D$ = 100 nm and $D$ = 1 mm.
Here the $D$ values are larger than the  calculated $D_c$ in Fig. 1(f), thus three-stage transition is possible at high field.
In Fig. 3(f), we show another $k$ = 0 case by using the FE model of a two-dimensional electron gas (2DEG) based on a AlGaAs/GaAs quantum-well-structure (QWS) \cite{inproceedings}.
For this case, we have $D_c \approx {2.4} \times 10^{-3}$ = 2.4 mm.
In contrary to Fig. 3(c), we have the only FE expected in the sub-critical regime for small $D < D_c$ (see Fig. 1(e) and Table I).

\textcolor{blue}{\textbf{\emph{Conclusion.}}} 
This paper shows that the conventional transition from the electron field emission (FE) model at low voltage to space charge limited current (SCLC) at high voltage is no longer valid when the FE does not follow the well-known Fowler-Nordheim (FN) law. 
For non-FN based or unconventional FE emission with arbitrary $k$, our model predicts a critical value of $k_c \equiv 3/2$ for which some peculiar transitions will appear in three different regimes ($k > k_c$, $k = k_c$ and $k < k_c$) for different gap spacing (as summarized in Fig. 1 and Table I).
Our model is formulated in a normalized form that is applicable over a wide range of materials and parameters. The unified picture of the emission physics from FE to SCLC has not been reported to our best knowledge. 
Thus, this paper offers a theoretical foundation for modelling different unconventional field emission as high current electron sources, which are important for many applications in high-power microwave sources, compact vacuum electronics, beam physics and plasma.

\begin{acknowledgments}
\textcolor{blue}{\textbf{\emph{Acknowledgments.}}} We thank B. Lepetit for insightful discussions. This work is supported by Singapore Ministry of Education Academic Research Fund (MOE AcRF) Tier 2 grant (2018-T2-1-007), A*STAR AME IRG grant (RGAST2007) and SUTD Startup Grant (SRT3CI21163). 
\end{acknowledgments}


\providecommand{\noopsort}[1]{}\providecommand{\singleletter}[1]{#1}%

\end{document}